# Exploring Post-Quantum Cryptographic Schemes for TLS in 5G Nb-IoT: Feasibility and Recommendations

**Kadir SABANCI**
Marquette University - Department of Computer Science
kadir.sabanci@marquette.edu

**Mumin CEBE**
Marquette University - Department of Computer Science
mumin.cebe@marquette.edu

**Abstract**
Narrowband Internet of Things (NB-IoT) is a wireless communication technology that enables a wide range of applications, from smart cities to industrial automation. As a part of the 5G extension, NB-IoT promises to connect billions of devices with low-power and low-cost requirements. However, with the advent of quantum computers, the incoming NB-IoT era is already under threat by these devices, which might break the conventional cryptographic algorithms that can be adapted to secure NB-IoT devices on large scale. In this context, we investigate the feasibility of using post-quantum key exchange and signature algorithms for securing NB-IoT applications. We develop a realistic ns-3 environment to represent the characteristics of NB-IoT networks and analyze the usage of post-quantum algorithms to secure communication. Our findings suggest that using NIST-selected post-quantum key-exchange protocol Kyber does not introduce significant overhead, but post-quantum signature schemes can result in impractical latency times and lower throughputs.

## Introduction

IoT technology has the potential to connect billions of smart devices in diverse locations, from urban centers to rural areas and even in space (Kua, Arora, Loke, Fernando, & Ranaweera, 2021). With this, various IoT applications are emerging in the transportation, healthcare, automation, and smart city industries, known as Industrial IoT (IIoT) applications (Sisinni, Saifullah, Han, Jennehag, & Gidlund, 2018). These applications are envisioned to use 5G/6G technologies that are designed to handle communication between IoT devices. An example of this is Narrow Band - IoT (NB-IoT) technology which can work with both 4G and 5G and offers greater coverage with improved signal penetration (Valecce, Petruzzi, Strazzella, & Grieco, 2020) at the cost of reduced bandwidth and therefore lower bitrates. However, the expansion in coverage also means connecting more IoT devices, which can put a lot of pressure on IoT applications due to increased communication traffic. Furthermore, the increased traffic also raises concerns about security management, as the security protocols can create additional overhead on traffic (Zhang, Wang, & Zhou, 2019).

Therefore, there are ongoing efforts to manage the security keys and certificates needed to serve millions of IoT devices, as reported in various research studies such as Narayanan et al. (2018), De Ree et al. (2019), Huang et al. (2021), and Hewa et al. (2020). Alongside these efforts, there are also ongoing studies to completely overhaul the security infrastructure, making communication protocols quantum resistant. However, quantum-resistant security schemes come at the cost of increased communication and computation overhead. With fast-pacing developments in quantum computing in recent years, it is believed that the first generation of industrial quantum computers will soon be developed (Gambetta, 2020). This has raised concerns about the implications of these computers for the security of IoT systems too. On the one hand, with the increased data traffic, there will be little bandwidth left to perform security management operations such as key exchange that is required for secure communications. On the other hand, migrating the post-quantum security protocols may add additional overhead, potentially causing delays that could compromise the application requirements.

This paper addresses the research question of the feasibility of deploying post-quantum cryptography on NB-IoT networks, considering the contradicting aspects of both technologies. Thus, in this study, we aim to evaluate recent post-quantum cryptography (PQC) algorithms as recommended by NIST when employedin supporting key management and authentication for the Transport Layer Security (TLS) on the NB-IoT network. The main objectives of this research are: 1) To evaluate the overhead of the substitution of the post-quantum key-exchange scheme with the Diffie-Hellman key-exchange in TLS handshake 2) To evaluate the overhead of post-quantum signature schemes to provide authentication in TLS handshake and the overhead of them to provide certificate verification. To achieve this, we use the ns-3 simulator to model and build a realistic simulation infrastructure that carries the NB-IoT characteristics as a low-bandwidth communication infrastructure. Then, we utilize the realistic NB-IoT setup to compare different conventional TLS cipher suites that use conventional keys and signatures with various post-quantum schemes that use post-quantum keys and signatures.

Currently, there are limited studies that evaluate the performance of post-quantum algorithms. The first notable study is conducted by (Sikeridis, Kampanakis, & Devetsikiotis, 2020), which compares the performance of some NIST post-quantum signature candidates selected by authors, on a broadband connection between a cloud server and a client. In 2021, (Paul, Schick, & Seedorf, 2021) modifies TLS 1.2 to evaluate the effects of the post-quantum key-exchange algorithm Kyber with post-quantum signature scheme Sphincs+ on a resource-constrained device based on TPM, with a focus on measuring computational overhead of the two post-quantum algorithms in TLS. Similarly, (Paquin, Stebila, & Tamvada) evaluates post-quantum schemes on TLS by assessing the overhead of hybrid schemes that use post-quantum and conventional key exchange at the same time during TLS handshake. The work aims to measure the overhead of hybrid schemes while establishing a TLS connection on a regular network. Finally, in 2023 (Kampanakis & Lepoin, 2023) conducts a benchmarking study of post-quantum schemes for QUIC protocol, which is a light-weight version of TLS protocol and evaluates the overhead of Dilithium-2 and Dilithium-3 certificate chains in cloud environment.

To the best of our knowledge, this study is the first comprehensive work that evaluates the effects of different post-quantum schemes in TLS on an NB-IoT network. In addition, we assess the overhead of post-quantum key-exchange (KEM) and post-quantum signatures by explicitly comparing the former with conventional key-exchange ECDHE and comparing the latter with conventional signature schemes ECDSAand RSA. By doing so, we shed light on performance differences of different post-quantum adaptation scenarios, such as separately evaluating the effect of post-quantum KEM and signatures on TLS. This provides valuable insights into the feasibility of adopting post-quantum schemes in various scenarios and highlights the areas where further considerations are needed according to the NB-IoT setup and the end- device density.

| Abbreviation | Definition |
| --- | --- |
| NB-IoT | Narrowband Internet of Things |
| PQC | Post-Quantum Cryptography |
| TLS | Transport Layer Security |
| KEM | Key Exchange Mechanism |
| ECDHE | Elliptic Curve Diffie-Hellman Ephemeral key exchange |
| MCL | Maximum Coupling Loss |
| LPWAN | Low Power Wide Area Network |
| RSA | Rivest–Shamir–Adleman Digital Signature Algorithm |
| ECDSA | Elliptic Curve Digital Signature Algorithm |

This paper is organized as follows; Preliminaries section makes an introduction to NB-IoT technology, TLS, and NIST-selected post-quantum schemes, whereas the next section makes an overview of the related works. In the Evaluations section we introduce the simulation environment and discuss the experiment results, in the final section we conclude the paper.

## Preliminaries

### NB-IOT

NB-IoT, or Narrowband Internet of Things, is a low-power wide-area network (LPWAN) protocol designed to improve coverage and reduce energy consumption while serving large number of Internet of Things (IoT) devices, such as sensors, smart meters, and tracking devices. It was officially standardized in 2016 with the 3GPP Release 13. Coverage enhancement modes that provide up to 20 dB improved maximum coupling loss (MCL) compared to GSM, which makes deep indoor signal penetration possible, and introduction of extended discontinuous reception mode (eDRX) which further improves the battery lifetime are some important updates brought about by this technology (Mwakwata, et al., 2019).

NB-IoT is designed with scalability in mind, enabling it to serve thousands of IoT devices from a single cell tower by sharing a single LTE resource block among many. To achieve this, NB-IoT uses a narrowband approach, employing a 180 kHz LTE Physical Resource Block (PRB) for both uplink and downlink channels. This narrowband approach is achieved by dividing the band into 12 subcarriers, each with 15kHz for user equipment which is significantly lower than the LTE bandwidth of 20 MHz (3GPP, 3GPP TS 36.104 version 13.4.0 Release 13, 2016). However, as NB-IoT uplink and downlink channels use different Transport Block sizes and different scheduling configurations, downlink bitrate is significantly lower than uplink (TechPlayon, 2020). Narrowband channel usage also enables NB-IoT to be deployed in various modes, including in-band on a single resource block of the LTE band, in a vacant spectrum between different LTE bands (guard band), and in a standalone band allocated in the GSM spectrum (Kousias, et al., 2020).

Furthermore, NB-IoT offers three coverage enhancement modes: CE0, CE1, and CE2. CE0 has the same MCL as GSM, which is 144 dB, whereas CE1 and CE2 provide 154 dB and 164 dB MCL respectively (3GPP, 3GPP TR 45.050 version 13.1.0 - Release 13, 2016). Increased MCL is achieved through repetitions. Nb-IoT can use up to 2048 repetitions in the uplink and 128 in the downlink, which is left to the network operator's preference to satisfy the best coverage while keeping connected devices battery efficient (Tabbane, 2018). In addition to these technical capabilities in the physical layer, NB-IoT also allows the integration of half-duplex NB-IoT terminals (Rohde&Schwarz), which significantly help to reduce the overall cost of IoT device deployments since these terminals are cheaper than full-duplex terminals. This advantage is particularly significant for IoT applications that involve infrequent, low-bandwidth communication, where the latency introduced by half-duplex communication may not be a major concern.

### TLS Protocol

TLS is a cryptographic protocol that provides security to TCP/IP communications. It is widely used for securing web traffic while also being used with many other communication protocols like email and DNS (Internet Society, n.d.). TLS is categorized as a hybrid protocol as it uses symmetric encryption together with asymmetric encryption. Symmetric encryption (TLS Record Protocol) is used for providing confidentiality to messages, whereas asymmetric encryption (TLS Handshake Protocol) is used to decide security parameters, authentication of communicating parties through digital certificates, and creation of session keys (Sullivan, 2018).

TLS 1.3 is the most recent version of the protocol which was standardized by IETF in 2018. With this version, the TLS handshake protocol is redesigned to improve security by starting data encryption earlier and decreasing latency by reducing the number of round trip times (RTT) of TLS 1.2 from 3-RTT to 1.5-RTT. This newly introduced key exchange mechanism is one of the most significant changes in TLS 1.3. The message flows in Figure 1 represent a server-only authentication handshake in public key infrastructure (PKI) settings with RSA or ECC signatures.

TLS implements two important security functions: key exchange and authentication. The client initiates the process by sending a "ClientHello" message, which includes the cryptographic parameters and a nonce. Additionally, as TLS 1.3 uses the Diffie-Hellman (DH) Key Exchange scheme, the client also sends a newly generated DH "KeyShare". The server responds with a "ServerHello" message, which contains its selected cryptographic parameters and a server nonce. The server also sends its own freshly generated DH

"KeyShare" and related "Extensions". To be authenticated, the server includes its "Certificate" (certificate chain) and a signature in the "CertificateVerify" message. The server's "Finished" message is a Message Authentication Code (MAC) computed over the entire handshake, ensuring the integrity of the shared key.

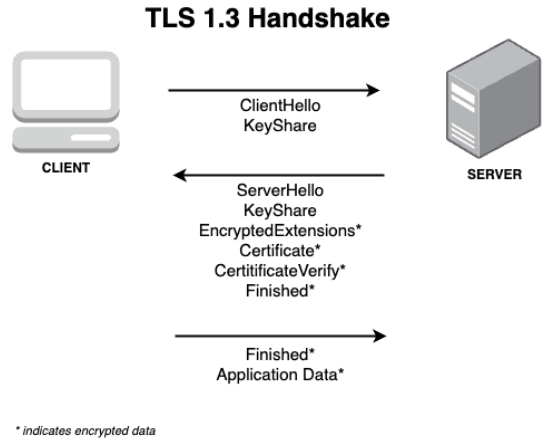

*Figure 1 -TLS 1.3 1.5-RTT Handshake*

In the case where the client wants to authenticate itself using a client certificate, the client sends its "Certificate" (certificate chain) and signature in "CertificateVerify" message along with "Finished" message and "ApplicationData". The client's "Finished" message is also a MAC over the entire handshake, providing both integrity and key confirmation. If client authentication is not required, such as in daily web traffic, the "Certificate" and "CertificateVerify" parts of the handshake message will be empty.

## *Post-Quantum Cryptography*

In recent years, encryption landscape is going through some drastic changes because of considerable development efforts put on quantum computers. Quantum computers are theoretically shown to be capable of solving a set of mathematically hard problems that lay the foundations of modern cryptography. In 1995, Peter W. Shor published his paper which proposes a quantum algorithm to solve prime factorization and discrete logarithm problems in polynomial time (Shor, 1997), which ultimately means that today's public key cryptography namely RSA or ECC will be broken once the quantum computing is available at scale.

NIST Post-Quantum Cryptography (PQC) standardization process started in 2017 for selecting quantum-resistant public-key cryptographic algorithms to replace the current vulnerable algorithms. Initial submissions to the NIST's process came from these 5 different families of encryption (NIST - Information Technology Laboratory, 2017):

**Lattice-Based:** Idea is basically creating problems on multidimensional lattices which are hard to solve even for quantum computers (Alwen, 2018). Some of the hard problems are Learning-with-Errors (LWE), Learning-with-Rounding (LWR), and Short Integer Solution (SIS) (Peikert, 2022). Lattice-based schemes are high performing in key generation, encapsulation, and decapsulation, and have reasonable public key and signature sizes (Buchanan B. , 2022). Crystals- Kyber (KEM), SABER (KEM), NTRU-Prime (KEM), Frodo (KEM), Falcon (DSA), and Crystals-Dilithium (DSA) are lattice-based schemes listed in NIST Round-3 either as a finalist or alternate candidates.

**Code-Based:** This family is based on error-correcting codes and their usability in encryption. When a random linear code is used for encryption, there is no efficient decoder, whereas if a good code is used, there exists a decoder that can be used for decryption (Sendrier, 2017). Classic McEliece (KEM), BIKE(KEM), and HQC(KEM) algorithms from this family are listed in the Round-3 qualifiers.

**Isogeny-Based:** This family is based on isogenies between two elliptic curves, which can be represented by polynomials where an addition operation in one curve would yield the same result when computed with the corresponding images on the second curve (ISARA Corporation, 2019). SIKE (KEM) is listed in Round-3 as an alternate candidate but was acknowledged as being insecure by the submitters, and there are no candidates left for this category in the standardization process (SIKE Team, 2022).

**Hash-Based:** The idea is to employ hash functions to generate one-time or many-time signatures such as Lamport or Merkle signatures which are believed to be safe against quantum attacks (Buchanan W. J., Hashed-based signatures, 2023). Sphincs+ is a Round-3 alternate candidate which uses many few-time signature (FTS) keys from which a random key pair is chosen to sign a message (Daniel J. Bernstein, 2017). Public and private key sizes for Sphincs+ remain small but the signature size is relatively higher than most of the other Round-3 qualifiers (Buchanan W. J., PQC Digital Signature Speed Tests, 2023).

**Multivariate:** These are the schemes based on multivariate polynomials over finite fields. Decryption is based on the hidden structure of the polynomial, so inverting the polynomial is possible (Ding, n.d.). Round-3 finalist Rainbow (DSA) has small signatures but comes with large key pairs (Buchanan W. J., PQC Digital Signature Speed Tests, 2023).

On July 2022 NIST announced the first selected algorithms from Round-3, as Crystals-Kyber for Key Exchange Mechanism (KEM); Crystals-Dilithium, Falcon, and Sphincs+ for Digital Signature Algorithms (DSA). For Round-4 there were no DSA candidates left for consideration at the time of writing this paper and one of the most important aspects of the selected DSA's is their large signature sizes compared to algorithms in use today, see Table 1 (NIST - Information Technology Laboratory, n.d.).

| Signature Algorithm and Parameter | NIST Classical Security Level | Signature Size (bytes) | Public Key Size (bytes) |
|---|---|---|---|
| RSA-2048 | <1 | 256 | 256 |
| RSA-3072 | 1 | 384 | 384 |
| ECDSA-prime256v1 | 1 | 64 | 64 |
| SPHINCS-SHA256-128s-simple | 1 | 7856 | 32 |
| SPHINCS-SHA256-192s-simple | 3 | 16224 | 48 |
| SPHINCS-SHA256-256s-simple | 5 | 29792 | 64 |
| Falcon-512 | 1 | 690 | 897 |
| Falcon-1024 | 5 | 1330 | 1793 |
| Dilithium2 | 2 | 2420 | 1312 |
| Dilithium3 | 3 | 3293 | 1952 |
| Dilithium5 | 5 | 4595 | 2592 |

*Table 1 - Signature and Public Key Size Comparison of traditional and post-quantum signature algorithms (Buchanan W. J., PQC Digital Signature Speed Tests, 2023), (Barker, 2020)*

# Evaluations
## *Experimental Setup*
We created a simulation environment in ns-3 version 3.32 (nsam, 2023) which employs the LTE implementation from (Gebauer & Jörke, 2023) to mimic an NB-IoT network. The simulation program we created (Sabanci, 2022), is a TCP application that fulfills data exchange steps between a client/server as it happens in a real-world TLS 1.3 handshake. Data sizes to be transmitted in each step were calculated using *s_client* and *s_server* programs from OpenSSL fork (Open Quantum Safe Github, 2023) of the OQS project, which supports post-quantum algorithms. Certificate chains to be used for authentication purposes were also created using the same program suite. Using ns3, we generated networks with various sizes by randomly distributing different number of IoT devices within a circular area of a one-kilometer radius. The central point of the circle was occupied by an evolved Node B (eNB) device, acting as a base station to connect the IoT devices in the CE0 coverage area, reflecting the ideal connectivity conditions. We just allowed a single eNB channel with a single resource block (RB) of 180KHz, which limits at most 12 IoT devices to connect simultaneously. When the number of connected IoTs exceeds 12, the rest will have to wait for others to complete communication to transmit their data. We particularly limit the base station with just a single NB-IoT channel to observe the effects of channel saturation. However, in a real-world setup, one should expect the availability of multiple cells at the same location which means that each base station could handle multiple channels to support various numbers of NB-IoT devices.

*Baselines*

The purpose of this section is to demonstrate the impact of post-quantum schemes on the communication overhead of the TLS 1.3 handshake in an NB-IoT network, in comparison to other conventional security schemes. The most important contributor to the traffic overhead during TLS handshake is digital signatures, which come in 2 different forms:

a) the signatures carried in certificates; in a server-only authentication, the server typically sends at least two digital certificates such as its own certificate a.k.a server certificate, and certificate of the issuing/intermediate Certificate Authority (CA) in "Certificate" field of the handshake,

b) the signature in the "CertificateVerify" message which provides integrity of the handshake and authenticity of the server.

The digital signatures mentioned in a) and b) make selection of the signature algorithm an important decision for resulting TLS performance.

To evaluate the effects of post-quantum schemes at different stages of the handshake, we have formed the following scenarios:

1. As a benchmark, we use the Diffie-Hellman key-exchange algorithm with ECC and RSA signature schemes, where server certificates contain either RSA public key & RSA signature pairs or ECC public key and ECC signature pairs. Key sizes are chosen as 2048 bits for RSA and 256 bits for ECC as they are the most commonly used key sizes (Warburton & Vinberg, 2022).
2. To assess the impact of post-quantum key-exchange algorithms, we replaced the Diffie-Hellman Key-exchange in TLS with the post-quantum key exchange *kyber-512* (Kyber, n.d.).
3. To evaluate the effects of post-quantum authentication overhead and corresponding increased certificate sizes due to the increased key and signature sizes, we created server, intermediate, and root CA certificates that contain public keys and signatures from the NIST-selected Falcon, Dilithium, and Sphincs+ post-quantum schemes. For Falcon and Sphincs+ we chose falcon-512 and sphincsha256128fsimple variants respectively which come with the smallest public key and signature sizes, on the other hand, we used the *dilithium3* variant with Crystals-Dilithium which was the recommended parameter set to achieve AES-128 bit security by the creators of the scheme (Dilithium, n.d.). All signatures were generated using *OpenSSL* (Open Quantum Safe Github, 2023), with the same key variants for server, intermediate and root CAs.

*Assessing the Overhead of Post-Quantum Key-Exchange (KEM) on TLS-Handshake*

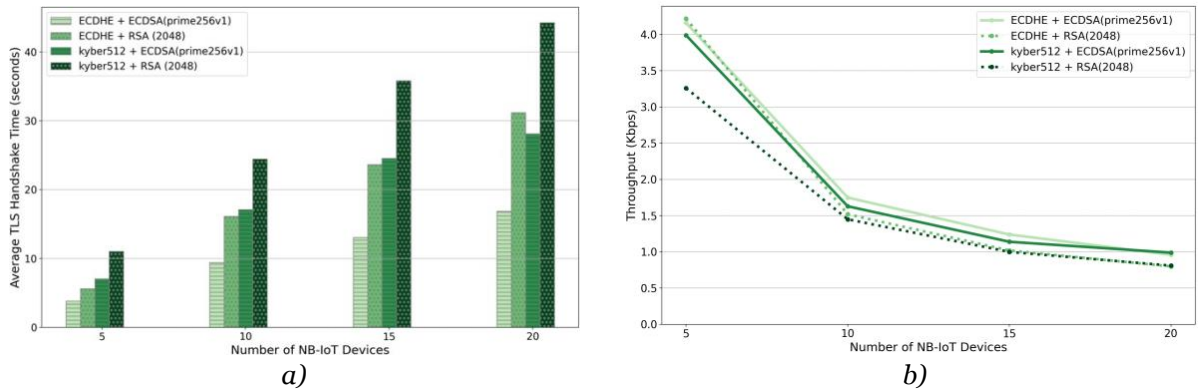

*Figure 2 - Effect of Key-exchange Algorithm on a) Average TLS Handshake Time, b) Throughput*

We first investigate the additional communication overhead of post-quantum key exchange (KEM) compared to conventional Elliptic-Curve Diffie Hellman Ephemeral (ECDHE) key exchange. To do so we substitute ECDHE with Kyber in two TLS cipher suites, namely TLS-**ECDHE**-RSA and TLS-**ECDHE**-ECDSA. The RSA and ECDSA components of these cipher suites represent the signature algorithms utilized. This allows us to assess the performance of Kyber across different signature suites too. Figure 3a illustrates the average completion time of the TLS handshake under varying number of IoT devices on a single NB-IoT channel. According to the experiment results shown in Figure 3a, replacing ECDHE with the post-quantum algorithm Kyber had a noticeable negative effect on average time to complete a TLS handshake. Specifically, when comparing ECDHE-RSA and Kyber-RSA, the average TLS-handshake time increased from 5.61 seconds to 11.02 seconds for just 5 NB-IoT devices. The results become even more dramatic when having 20 devices in the channel, where the average TLS-handshake time almost hits one minute. Similarly, when comparing ECDHE-ECDSA and Kyber-ECDSA, the average time increases from 3.85 seconds to 7.07 seconds for 5 devices. However, we observe that the increase is not as dramatic when using ECDSA as the signature algorithm instead of RSA and the TLS handshake takes under 30 seconds on average for 20 devices. This is because the small overhead ECDSA helps the NB-IoT network to better accommodate the overhead of post-quantum Kyber. In both cases, however, Kyber introduces a significant overhead, and this overhead can be directly attributed to the difference in key_share size between Kyber and ECDHE algorithms as shown in Table 2. The key_share size of Kyber in ClientHello message is 806 bytes, which is 768 bytes larger than the 38 bytes of ECDHE. The size increase is also observed in the ServerHello message which was 36 bytes for ECDHE and 772 bytes for Kyber. Overall, the substitution of key-exchange algorithms in TLS handshake adds an overhead of 1500 bytes which causes additional 50% overhead compared to ECDHE + RSA and 73% compared to ECDHE + ECDSA as can be deducted from Table 2. These findings highlight the impact of key-exchange algorithm selection on network traffic size, particularly in the context of post-quantum algorithms and their larger key sizes.

| KEM/Digital Signature | Key_share size in ClientHello (bytes) | Key_share size in ServerHello (bytes) | Total Traffic Exchanged During Handshake (bytes) |
|---|---|---|---|
| ECDHE + ECDSA (prime256v1) | 38 | 36 | 2048 |
| ECDHE + RSA (2048) | 38 | 36 | 3026 |
| kyber512 + ECDSA (prime256v1) | 806 | 772 | 3560 |
| kyber512 + RSA (2048) | 806 | 772 | 4522 |

*Table 2 - Traffic size comparison of key-exchange algorithms ECDHE vs Kyber*

We also investigate the throughput in our experiments and observe that NB-IoT provides a comparable throughput for Kyber against the ECDHE scheme. As shown in Figure 3b while throughput for all scenarios starts from roughly 4 kbps for 5 IoT devices, it gradually decreases to 1 kbps for 20 devices in ECDSA scenarios and 0.8 kbps for RSA scenarios. This observation indicates that the asymmetrical bitrate characteristic of NB-IoT networks favors uplink transmission rather than downlink, which helps tolerate overhead coming from bigger key_share of Kyber while not performing as well when transmitting bigger RSA signatures in the downlink direction.

### *Assessing the Overhead of Post-Quantum Signature Algorithms on TLS-Handshake*

As the second benchmark, we look into overhead created by post-quantum signature algorithms compared to ECDSA when post-quantum Kyber is used as the default key-exchange algorithm. Post-quantum signature algorithms coming with big signature and public key sizes produced huge latency spikes in our tests. The most light-weight post-quantum signature scheme tested is Falcon as shown in Table 3, thus it gives the best results with below 16 seconds on average to complete TLS handshake for 5 devices, and about 50 seconds for 20 devices. Dilithium and Sphincs+ remain above 40 and 70 seconds respectively even having just 5 devices in the channel. This outcome is understandable referring to the total exchanged traffic sizes of these two algorithms which are 16997 and 26830 bytes respectively.

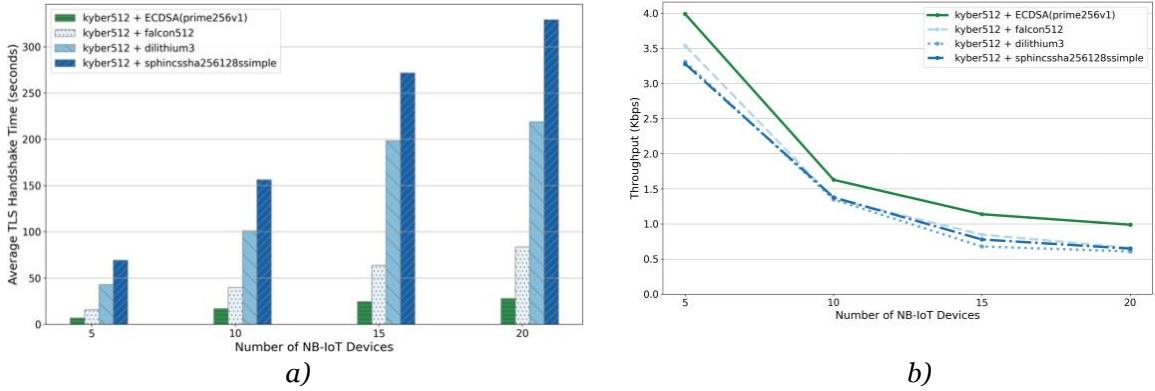

*Figure 3 - Effect of Signature Algorithm on a) Average TLS Handshake Time, b) Throughput*

Initial throughput results for 5 devices are about 3.5 kbps which is slightly lower than ECDSA/RSA scenarios, but with increased number of devices, results get even lower. Compared to ECDSA, we got 28% lower throughput in 10, 15, and 20 devices with post-quantum authentication schemes, ending up with 0.6 kbps of throughput, which means NB-IoT is adversely affected by the large traffic overhead introduced by new post-quantum signature schemes.

| KEM/Digital Signature | ICA + Server Certificate Size (bytes) | Signature Size (bytes) | Total Traffic Exchanged During Handshake (bytes) |
|---|---|---|---|
| kyber512 + ECDSA (prime256v1) | 592 | 64 | 3560 |
| kyber512 + falcon512 | 3404 | 690 | 6965 |
| kyber512 + dilithium3 | 10824 | 3293 | 16997 |
| kyber512 + sphincsha256128fsimple | 16070 | 7856 | 26830 |

*Table 3 - Certificate and signature overheads of current vs post-quantum signature algorithms with Kyber key-exchange*

As a result, substituting conventional signature schemes with post-quantum signature schemes in TLS handshake has a significant effect on performance compared to just substituting the conventional key-exchange scheme with post-quantum key-exchange scheme Kyber. Falcon introduces an overhead of nearly 5 folds compared to ECDHE/ECDSA and 2.5 folds compared to ECDHE/RSA. However, it is still the best performant among post-quantum signature schemes, and 15 seconds latency with 5 devices, makes it the feasible candidate for post-quantum signature migration if it is used in low-density NB-IoT networks. The results show us Dilithium and Sphincs+ might be impractical to use for most applications that run on IoT networks.

## Discussions & Conclusions

This study examined the feasibility of using post-quantum cryptographic schemes in TLS to ensure the security of NB-IoT applications, focusing on use cases in open fields such as agriculture, smart grids, and pipelines. Based on the experiment results, we can conclude that NB-IoT applications can switch to post-quantum key-exchange KYBER with ECDSA in TLS, with an acceptable performance loss. However, using post-quantum signature schemes would lead to impractical TLS handshake times and lower throughputs. If the adaptation of post-quantum signatures is necessary, the most feasible option is Falcon with ring degree 512.

In light of these findings, we recommend that migrating to Kyber is a viable option to prevent potential "store-now, decrypt later" type attacks by breaking ECDHE keys with quantum computers in the future (Crockett, Paquin, & Stebila, 2019). However, since currently there is no evident threat, it is better to continue using a conventional signature scheme in TLS authentication and certificates until a quantum computer powerful enough to break the security of ECDSA is developed.

However, our study has some limitations:
- The open field assumption in our ns-3 experiments may not accurately represent use cases for industrial IoT settings (IIoT) with obstacles, impacting the generalizability of our findings.
- Our experiment was conducted in the CE0 coverage area, which may not capture the full range of network conditions.

Future research should address these limitations by exploring more complex environments, including CE1 and CE2 coverage setups, and dense urban areas to cover IIoT use cases.